\documentclass{emulateapj}

\def\mh{M_{\rm BH}}
\def\mb{M_{\rm b}}
\def\vk{V_{\rm k}}
\def\vesc{V_{\rm esc}}

\def\rk{r_{\rm k}}
\def\rinfl{r_\bullet}
\def\vesc{V_{\rm esc}}

\def\kms{km s$^{-1}$}

\newcommand{\gap}{\;\rlap{\lower 2.5pt \hbox{$\sim$}}\raise 1.5pt\hbox{$>$}\;}
\newcommand{\lap}{\;\rlap{\lower 2.5pt \hbox{$\sim$}}\raise 1.5pt\hbox{$<$}\;}
\newcommand{\beq}{\begin{equation}}
\newcommand{\eeq}{\end{equation}}
\newcommand{\msun}{M_\odot}

\shorttitle{Tidal Flares from Recoiling Black Holes}
\shortauthors{Komossa \& Merritt}

\begin{document}

\title{Tidal Disruption Flares from Recoiling Supermassive Black Holes} 

\author{Stefanie Komossa} 
\affil{Max-Planck-Institut f\"ur extraterrestrische Physik,
Postfach 1312, 85741 Garching, Germany; skomossa@mpe.mpg.de}
\author{David Merritt}
\affil{Center for Computational Relativity and Gravitation and
Department of Physics, Rochester Institute of Technology,
Rochester, NY 14623, USA; merritt@astro.rit.edu}

\begin{abstract}
Supermassive black holes ejected from galaxy nuclei by gravitational wave
recoil will carry a retinue of bound stars, 
even in the absence of an accretion disk.
We discuss the observable signatures related to these stars,
with an emphasis on electromagnetic flares from stars which
are tidally disrupted by the black hole.
We calculate disruption rates for the bound, and the unbound,
stars. The rates are smaller than, but comparable to, rates for non-recoiling
black holes.
A key observational consequence is the existence
of powerful off-nuclear and intergalactic X-ray
flares.
We also discuss other observable signatures associated with the
bound stars, including episodic X-ray emission from accretion due to stellar mass
loss, intergalactic supernovae, and feedback trails.
\end{abstract}

\keywords{galaxies: active -- galaxies: evolution -- galaxies: X-rays}

\section{Introduction}

Gravitational waves emitted during gravitational collapse
carry away linear momentum,
causing the center of mass of the collapsing object to recoil 
(Peres 1962, Bekenstein 1973).
In the case of binary black holes, 
coalescence can be accompanied by kicks as large as $\sim 200$ \kms~if
the holes are nonspinning prior to coalescence,
increasing to $\sim 4000$ km\,s$^{-1}$ 
in the case of maximally spinning, equal-mass holes with
the optimum orientations
(e.g., Campanelli et al. 2007; Gonz{\'a}lez et al. 2007;
Herrmann et al. 2007; Koppitz et al. 2007; Baker et al. 2008; 
Schnittman et al. 2008).

Recently, the first compelling candidate for a recoiling SMBH
was discovered (Komossa et al. 2008).
The quasar SDSSJ092712.65+294344.0 at $z=0.7$  exhibits two separate sets of unusual emission lines: 
a set of broad emission lines, presumably associated with gas
bound to a SMBH, and a second set of atypically narrow emission lines.
The broad lines are 
shifted by $\sim 2650$ \kms~relative to the narrow lines.
This is one of the key predicted signatures of a recoiling SMBH,
which would carry with it the gas from the broad line
region while leaving the bulk of the narrow line gas
behind (Merritt et al. 2006; Bonning et al. 2007).

Gas near to the SMBH is expected to respond to the kicks
in a number of other, potentially observable ways.
Accretion activity may be temporarily interrupted during the final
phases of binary coalescence
but would re-start thereafter with a certain time delay
(Liu et al. 2003; Milosavljevi{\'c} \& Phinney 2005).
The recoiling SMBH would then appear, temporarily, as a ``quasar'' that 
is spatially offset from the core of its host galaxy 
(Madau \& Quataert 2004, Loeb 2007).
UV (Lippai et al. 2008), soft X-ray 
(Shields \& Bonning 2008) and IR flaring (Schnittman \& Krolik 2008)
could result from
shocks in the accretion disk surrounding the SMBH
just after recoil, or when the inner disk reforms. 

All of these detection methods require the presence of an 
accretion disk around the coalescing SMBHs, 
and the duration of the predicted signal is limited 
to the time it would take the bound gas to be accreted,
or much less.
But such gas may be lacking in mergers of early type galaxies,
and even in gas-rich mergers, inspiral of the binary SMBH may
stall long enough that the nearby gas is mostly depleted.
How can we observe such recoils ?  
Even in the absence of a gaseous accretion disk,
a recoiling SMBH will always be accompanied by a retinue
of tightly bound stars (Gualandris \& Merritt 2008), 
and a fraction of these stars 
will eventually undergo tidal disruption. 
Stellar tidal disruptions (Luminet 1989) have been observed in the form 
of giant-amplitude, luminous X-ray flares 
(Komossa \& Bade 1999; Halpern et al. 2004; Komossa et al. 2004).
In the case of a recoiling SMBH, these flares will appear off-nuclear. 
In this {\it Letter} we present 
tidal disruption of stars
as one of the most universal observational
signatures of recoiling black holes, and discuss other 
signatures related to bound stars. 

\section{Disruption Rate of Bound Stars}

A recoiling SMBH carries with it a cloud of stars on bound orbits.
We first compute the mass of this cloud
(expressed as a fraction $f_b$ of the SMBH's mass),
then we compute the rate at which the stars would be 
scattered into the SMBH's tidal disruption sphere.

Figure~1a shows the steady-state distribution of bound stars following an
instantaneous kick of magnitude $\vk$, assuming an initial, 
power-law density profile $\rho=\rho_0 (r/r_0)^{-\gamma}$ around the SMBH.
\footnote{The bulk of the recoil velocity is imparted to the coalesced SMBH 
in a time $\sim G\mh/c^3$ \citep[e.g. Fig.~1 of][]{campanelli07}
and so the kick is essentially instantaneous as seen by the stars.}
Stars initially at distances 
$r\gap\rk\equiv G\mh/\vk^2$ from the SMBH
will be unbound after the kick. For $\vk\gap 0.4\vesc$
i.e. large enough to remove the SMBH from the galaxy core
(Gualandris \& Merritt 2008; $\vesc$ is the escape velocity),
$\rk$ is small compared to the SMBH influence radius 
and stars that follow the SMBH will be moving on essentially Keplerian orbits 
both before and after the kick.
The final distribution can be computed uniquely from the initial 
distribution by randomizing the orbital phases of the stars
that remain bound. 
Beyond $\sim \rk$ the cloud consists of an elongated, 
$\rho\sim r^{-4}$ envelope containing
stars that were pushed into nearly unbound orbits by the kick
(Fig.~1b).

The bound mass must scale as 
$\mb\sim \rho(\rk)\rk^3\approx\rho_0r_0^3(G\mh/r_0\vk^2)^{3-\gamma}$.
Choosing for $r_0$ the (pre-kick) influence radius $\rinfl$ of the SMBH,
defined as the radius containing a mass in stars equal to twice $\mh$,
this becomes
\beq
f_b\equiv {\mb\over\mh} = F(\gamma)
\left({G\mh\over \rinfl\vk^2}\right)^{3-\gamma};
\label{eq:mb1}
\eeq
we find $F(\gamma)\approx 11.6\gamma^{-1.75}$,
$0.5\lap\gamma\lap 2.5$.
An alternative form is
$f_{b,-3} = G(\gamma)M_{\bullet,7}^{3-\gamma}$
$r_{\rm infl,1}^{\gamma-3}
V_{\rm esc,3}^{2(\gamma-3)}
(\vk/\vesc)^{2(\gamma-3)}$
where 
$f_{b,-3} \equiv f_b/10^{-3}$,
$M_{\bullet,7}\equiv \mh/10^7\msun$,
$r_{\rm infl,1}\equiv \rinfl/10$ pc,
$V_{\rm esc,3}\equiv\vesc/10^3$ km s$^{-1}$,
and $G(\gamma) = (0.048,0.22,1.6,15)$
for $\gamma=(0.5,1,1.5,2)$.
For kick velocities in the range of interest,
equation~(\ref{eq:mb1}) predicts bound masses of order $1\%$
of the SMBH mass, falling as $\vk^{-2(3-\gamma)}$.
The stars around a recoiling SMBH would 
resemble a globular cluster in total luminosity,
but with a much greater velocity dispersion due to the
large binding mass $\mh$; observationally they might look
like ultracompact dwarf galaxies (e.g., Drinkwater et al. 2003).

The appropriate value for $\gamma$ is the density slope
just prior to the kick, and after the binary SMBH has completed
its inspiral; $\rinfl$ is likewise defined at this time.
Slow inspiral of the SMBHs in a large, low-density galaxy
produces a flat core, $\gamma\approx 0.5$ (e.g. Merritt 2006);
however it is not clear whether binary SMBHs in such nuclei
can overcome the ``final parsec problem'' (Milosavljevi{\'c} \& Merritt 2003) and coalesce.
In lower-mass galaxies, $M_{\rm gal}\lap 10^{10}\msun$,
binary evolution can continue to coalescence in $\le 10$ Gyr
via collisional loss-cone repopulation; in this case
the pre-kick density profile would have $\gamma\sim 1.75$
(Merritt et al. 2007).
Rapid, gas-driven inspiral would tend to preserve the
initial density profile and might even steepen it via star formation.
Henceforth we take $\gamma=1$ as a fiducial value. 

In standard loss-cone theory, stars
are scattered into the tidal disruption sphere 
$r\le r_t$ at a rate
$\sim N(r)/\left[t_r(r)\ln(r/r_t)\right]$ where $N(r)$ is the number of
stars within $r$ and
$t_r(r)$ is the local (non-resonant) relaxation 
time.\footnote{The logarithmic factor is associated with diffusive
loss-cone repopulation which is appropriate in the ``empty
loss cone'' regime near a SMBH (e.g. Lightman \& Shapiro 1977).}
The total disruption rate from the cloud of bound stars 
would be
\beq
\dot N_{NR} \approx C_{NR}(\gamma) {\ln\Lambda\over\ln(\rk/r_t)}
\left({\vk\over\rk}\right) f_b^2
\label{eq:ndot1}
\eeq
with $\ln\Lambda\approx \ln(\mh/m_\star)$ the Coulomb logarithm;
the ratio of logarithmic terms is of order unity.
Equation~(\ref{eq:ndot1}) assumes that gravitational encounters
are uncorrelated;
however, near the SMBH, orbits are slowly precessing Keplerian ellipses
and ``encounters'' are highly correlated (Rauch \& Tremaine 1996),
shortening the effective, angular momentum relaxation time by a factor 
$\sim m_\star N(r)/\mh$
(Hopman \& Alexander 2006).
The contribution of this ``resonant relaxation'' to tidal
disruption rates of SMBHs embedded in galaxies is small
since most of the disrupted stars come from orbits with $r\approx\rinfl$.
However in our  case, stars beyond $\sim \rk\ll\rinfl$ were removed 
by the kick and loss-cone repopulation is {\it dominated} by resonant 
scattering, yielding
\beq
\dot N_{RR} \approx C_{RR}(\gamma) 
{\ln\Lambda\over\ln(\rk/r_t)}
\left({\vk\over\rk}\right) f_b,
\label{eq:RR}
\eeq
i.e. $N_{RR}\gg N_{NR}$ for $f_b\ll 1$.

The coefficient $C_{RR}$ in equation~(\ref{eq:RR}) is poorly
determined (Rauch \& Ingalls 1998; Hopman \& Alexander 2006).
Because the number $\mb/m_\star$ of stars remaining bound to a 
recoiling SMBH is relatively small, tidal disruption rates in this regime can 
be computed directly via sufficiently accurate $N$-body integrations.
Figure~1c shows the results of a series of such experiments starting
from initial conditions generated from the steady-state (nonspherical
and anisotropic) distribution of Figure~1a, realized with $N=1.5\times 10^4$ 
particles and various values of $m_\star$.
We used the hybrid $N$-body code $\phi$GRAPEch \citep{harfst08}
running on the RIT GRAPE cluster \citep{harfst07}.
Stars were initially removed from the model if their periastra
fell below $10^{-4}\rk$, the assumed radius of the disruption
sphere; the model was then integrated forward and 
stars that approached the SMBH particle at distances $\le r_t$
were recorded and removed from the integration.
The $N$-body integrations confirmed the 
$\dot N\propto f_b$ dependence predicted by 
equation~(\ref{eq:RR}).
\footnote{We note that the various other conditions required for 
resonant relaxation to be present in these simulations
were satisfied,
i.e. the integration times were long compared with
orbital precession times and most of the stars were in the
diffusive, as opposed to pinhole, loss-cone regime.}
In the adopted units ($G=\mh=\vk=1$), the disruption rate 
is $\sim 0.15 f_b$; assuming that 
$\Lambda\approx \rk/(2Gm_\star/V^2)\approx \mh/m_\star$
in the simulations then implies $C_{RR}(\gamma=1)\approx 0.14$.

Using the derived value of $C_{RR}$, equation~(\ref{eq:RR}) predicts
for the disruption rate of bound stars ($\gamma=1$):
\beq
\dot N_{\rm b}\approx 6.5 \times 10^{-6} {\rm yr}^{-1}
M_{\bullet,7}^{-1} V_{k,3}^3 f_{b,-3} ;
\label{eq:rateRR}
\eeq
here and below, $\ln\Lambda/\ln(\rk/r_t)$ has been set to 2.
Combining equation~(\ref{eq:rateRR}) with equation~(\ref{eq:mb1}) for
the bound mass gives
\beq
\dot N_{\rm b}\approx 1.5 \times 10^{-6} {\rm yr}^{-2}
M_{\bullet,7}
r_{\rm infl,1}^{-2}
V_{k,3}^{-1}.
\label{eq:rateRRmb}
\eeq

Figure~2 (solid lines) plots $\dot N_{\rm b}$ for SMBHs ejected from the
centers of two representative galaxies, with masses
$4.5\times 10^{10}\msun$ and $1.5\times 10^9\msun$; the galaxies 
were modelled as Prugniel-Simien (1998) (i.e. de-projected Sersic)
spheroids with Sersic indices $(4,2.5)$.
How do the rates we estimate compare with flare rates of
non-recoiling SMBHs? 
The steady-state stellar disruption rates of SMBHs embedded
in galactic nuclei are predicted to be in the range
$10^{-5} {\rm yr}^{-1}\lap\dot N \lap 10^{-4} {\rm yr}^{-1}$ 
for $\mh\approx 10^7\msun$
(e.g. Wang \& Merritt 2004, Fig.~ 5b).
For $\vk\lap 10^3$\kms, rates derived here are roughly an
order of magnitude lower.
Were resonant relaxation not effective, this ratio would
be $\sim 10^3$ rather than $\sim 10$.

Ignoring changes in the shape of the density profile of the bound population
with time,
the disruption rate is predicted to drop off as 
$\dot N=\dot N(0)\exp(-t/\tau), \tau\approx 3.6G\mh^2/\vk^3m_\star$. 
This decay is included in the curves of Figure~2.
In fact, after $\sim 1$ (non-resonant) relaxation time, the density would
evolve to the Bahcall-Wolf $\rho\propto r^{-7/4}$ slope
and maintain this profile as its amplitude decayed;
this complication was ignored in the curves of Fig.~2.

We note that any post-merger galaxy 
would likely be highly inhomogeneous.  
Tidal disruption rates could be temporarily increased during
close encounters of the SMBH to a massive perturber, 
e.g. a giant molecular cloud, in much the same way that
``comet showers'' are triggered by near passage of the
solar system to a star \citep{hills81}.
The same is true for the rate of disruption of unbound stars
(\S 3) if the SMBH passes through a dense clump.

\section{Contribution of Unbound Stars}

Even a naked SMBH encounters stars as it traverses a galaxy
(Kapoor 1976).
Unbound stars are deflected into the tidal disruption sphere 
at an instantaneous rate
\begin{eqnarray}
\dot N_{\rm unb}
&\approx& 2\pi G\mh (\rho/m_\star) r_t V_\bullet^{-1} \nonumber \\
&\approx& 1.7\times 10^{-9} {\rm yr}^{-1} \tilde\rho(r/R_e)
\rho_{e,1}
M_{\bullet,7}^{4/3}
V_{_\bullet,3}^{-1}
R_{\star,11}
\label{eq:unbound}
\end{eqnarray}
where $\rho$ is the local (stellar) density,
$\tilde\rho\equiv\rho/\rho(R_e)$, $R_e$ is
the galaxy effective (projected half-light) radius,
$\rho_{e,1}\equiv\rho(R_e)/1\msun {\rm pc}^{-3}$, and
$R_{\star,11}\equiv R_\star/10^{11}$ cm;
$V_{\bullet}=V_\bullet(r)$ is the instantaneous SMBH velocity.
Figure~2 (dashed lines) shows $\dot N_{\rm unb}(r)$ 
in two representative galaxy models.
(The curves in Fig.~2 include the correction factor of 
Danby \& Camm (1957) that accounts for the decrease
in the capture rate when $V_\bullet\approx\sigma$,
with $\sigma$ the stellar velocity dispersion.)
In the larger of the two galaxy models considered, 
of order $10$ flaring events are predicted from both
bound and unbound stars for $\vk=10^3$ \kms during the 
time required for the SMBH to travel from the core to the 
half-mass radius; in the smaller galaxy,
$\dot N_{\rm unb}\ll \dot N_{\rm b}$.

\section{Observability}

We have shown that rates of stellar disruption by
recoiling SMBHs are interestingly high;
only moderately lower than those of SMBHs in the cores of galaxies. 
For a typical galaxy, we predict $\sim 20$ flares 
as the SMBH travels through the central parts of the galaxy
(50\% of these from bound stars, 
50\% from unbound stars), and $\sim 10^2-10^3$ more after the SMBH has
left the galaxy. 
Now, we discuss observational consequences. 

\subsection{Powerful off-nuclear X-ray flares and feedback trails}

Stellar tidal disruptions appear as luminous X-ray flares
which reach quasar-like 
luminosities and then
decline on month-year time scales (e.g. Komossa \& Bade 1999).
Tidal flares from recoiling SMBHs would look similar
but would occur off-nucleus,
and would be easily identified since no other known mechanism
produces quasar-like luminosities far from the nucleus. 

Future X-ray all-sky surveys like {\sl eROSITA} 
(Predehl et al. 2006) and
{\sl EXIST} (Grindlay et al. 2005) will be sensitive to these
types of events. While sky surveys are most suited to identifying the
candidates, follow-up observations with high
spatial resolution ($\sim0.5$\arcsec, Chandra) will
then allow the measurement  of spatial offsets from the galaxy core.
If the flare is associated with jet emission,
radio observations would greatly improve the positioning accuracy.
With future spectroscopic instrumentation, we will be
able to determine the line-of-sight recoil
velocity directly, if emission lines form in the temporary
accretion disk of the stellar debris. Such instrumentation will be
available aboard {\sl XEUS}, its NFI
reaching 3 eV resolution at 6 keV (Turner et al. 2007).
While amplitudes of variability are highest in the
soft X-rays, optical wide-field surveys 
like Pan-STARRS (Kaiser et al. 2002)
will also be sensitive to flare events.
  
Before the {\sl eROSITA} mission,
to be launched
in 2011, 
will start its all-sky survey, 
during the first
half year of its operation 
a deep survey of the nearest clusters
of galaxies will be performed (G. Hasinger, priv. com.).
This will provide an excellent opportunity
to detect the nearest flare events. 
Disruption rates of unbound stars are highest in the galaxy core (Fig. 2);
in order to identify these as off-nuclear events, high spatial
resolution is required
(at the distance of the Virgo cluster,
1$^{\prime\prime}$ corresponds to $\sim$100 pc).
The long time scales for damping of orbital oscillations 
(Gualandris \& Merritt 2008)
when $\vk\le V_{\rm esc}$ will
increase the chances of detecting an offset.
In addition, the abundance
of ``fossil'' intracluster recoiling SMBHs from past mergers 
(Volonteri 2007) would be
especially high in clusters.

While the brightest phases of a tidal disruption flare might only last
a few years or decades, interaction with the environment
would leave a feedback trail along the path of the SMBH. 
If radio jets are temporarily formed 
during the accretion phase
(Wong et al. 2007), they would create local cavities
in the ISM, similar to but smaller than
the X-ray cavities that have been observed in nearby
ellipticals (e.g., Biller et al. 2004)
and that have been suggested to be linked
to tidal disruption (Wang \& Hu 2005).
The bright flare will also excite emission lines in any surrounding
gas. Even though these lines would be faint given the short time span
of the energy input, low density gas would retain memory
of the flare for a long time 
(e.g., for a particle density $n$ = 10 cm$^{-3}$, the hydrogen recombination
time scale is on the order of 10000 yrs). 
This way, quasar-like emission-line ratios
could be  produced far  
away from the nucleus
of a galaxy.

In addition to stellar disruption and related observational
signatures, mass loss from evolving bound stars (e.g., Kudritzki \& Puls 2000) will 
provide episodic low-level fuelling of the SMBH,
causing repeated episodes of X-ray emission and variability. 

\subsection{Intergalactic flares and other signatures}

Tidal disruptions continue long after the recoiling
SMBH (with $\vk>V_{\rm esc}$) 
has left its host galaxy and wanders in intergalactic
space (Fig. 2). 
Such events will appear as luminous flares without host galaxies.
Furthermore, the bound stars will undergo
stellar evolution and will therefore ultimately produce intergalactic
planetary nebulae, and intergalactic supernovae (SNe) of type I
{\footnote{We note that high-mass
stars that produce type II SNe evolve quickly;
within $t < 100$ Myr, a SMBH moving at 1000 km s$^{-1}$ has
reached 100 kpc. Therefore, most of the type II SNe
would be produced when the SMBH has not yet escaped into
intergalactic space. 
These events might be identified from their large velocities
with respect to their host galaxy.}}. 
While the most massive recoiling SMBHs ($M > 10^8$ M$_{\odot}$) would generally
not disrupt solar-type stars,
they would still partially disrupt giant stars,
and they would also show
repeated phases of activity from fuelling due to stellar mass loss.
As such they might hide among the ``blank field sources''
with bright X-ray emission but no optical counterpart
(e.g. Cagnoni et al. 2002) if the absence of optical hosts
persists in deeper optical imaging.

After several Gyrs, stars will start turning into white dwarfs,
and SMBHs which left their galaxies long ago will have a cloud
of white dwarfs, neutron stars and stellar mass BHs bound to them. 
While the white dwarf tidal radius is inside the Schwarzschild radius
except for low ($\lap 10^{5-6} \msun$) BH masses,
Dearborn et al. (2005) have shown
that
relativistic compression causes white dwarfs to ignite and explode as SN at
distances
up to $100 R_s$ from a SMBH. Part of the disrupted
white dwarf will escape while the rest will eventually be accreted;
both the SN signal
and the accretion phase would allow us to detect such
fossil intergalactic SMBHs.

In summary, we have discussed observational signatures of
recoiling black holes related to the bound (and unbound) stellar
population. 
Finally, we emphasize that 
all these signals would {\it generically}
be associated with recoiling SMBHs, whether or not an accretion
disk is present initially, and that
they would continue episodically for a time of $\sim 10$ Gyr.

 \acknowledgments
 This work was supported by grants 
 AST-0420920 and AST-0437519 from the NSF and grant 
 NNG04GJ48G from NASA.

\clearpage

\begin{figure*}
\includegraphics[angle=-90.,width=1.\textwidth]{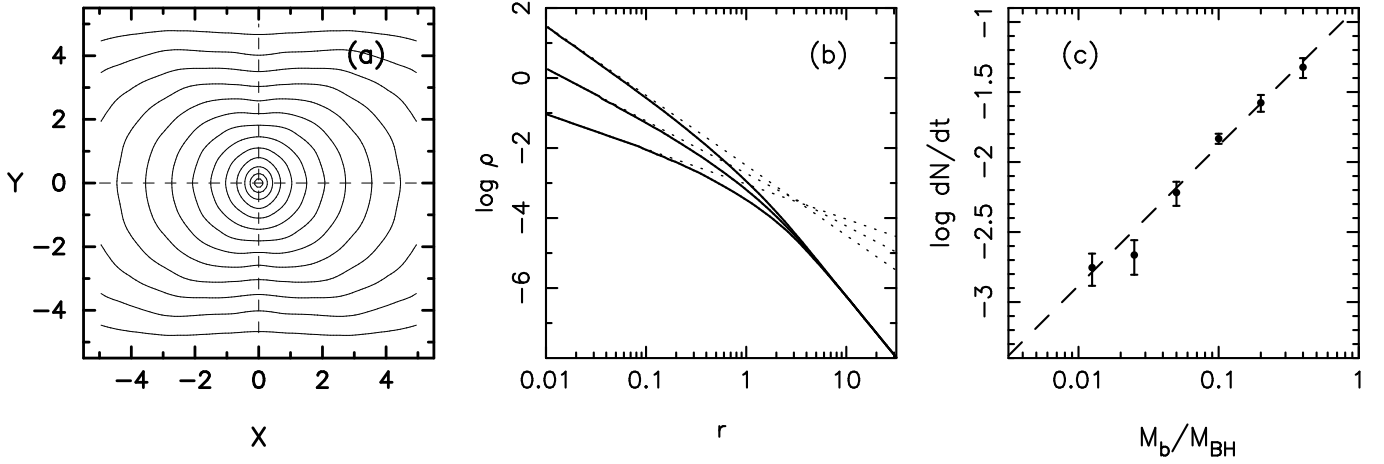}
\caption{(a) Contours of the projected density of a cloud
of bound stars around a kicked BH. $x$ and $y$ are spatial
scales and the unit of length is $G\mh/\vk^2$.
The initial density was $\rho\propto r^{-1}$ and the kick was in the 
$x$-direction.
Contours are separated by 0.17 in the log.
(b) Spherically-symmetrized density profiles of
stars around kicked BHs; the initial profiles 
-- $\rho\propto r^{-\gamma}, \gamma=(1,1.5,2)$ --
are indicated by the dotted lines. Density normalization is arbitrary. 
(c) Dependence of disruption rate on bound mass in a set of
$N$-body integrations with $N=1.5\times 10^4$ and various values
of the mass $M_b$ in stars bound to the BH.
$dN/dt$ is in $N$-body units, as defined in the text.
The dashed line has unit slope,
the expected dependence if feeding is dominated by 
resonant relaxation.
}
\end{figure*}

\begin{figure}
\includegraphics[width=1.0\textwidth]{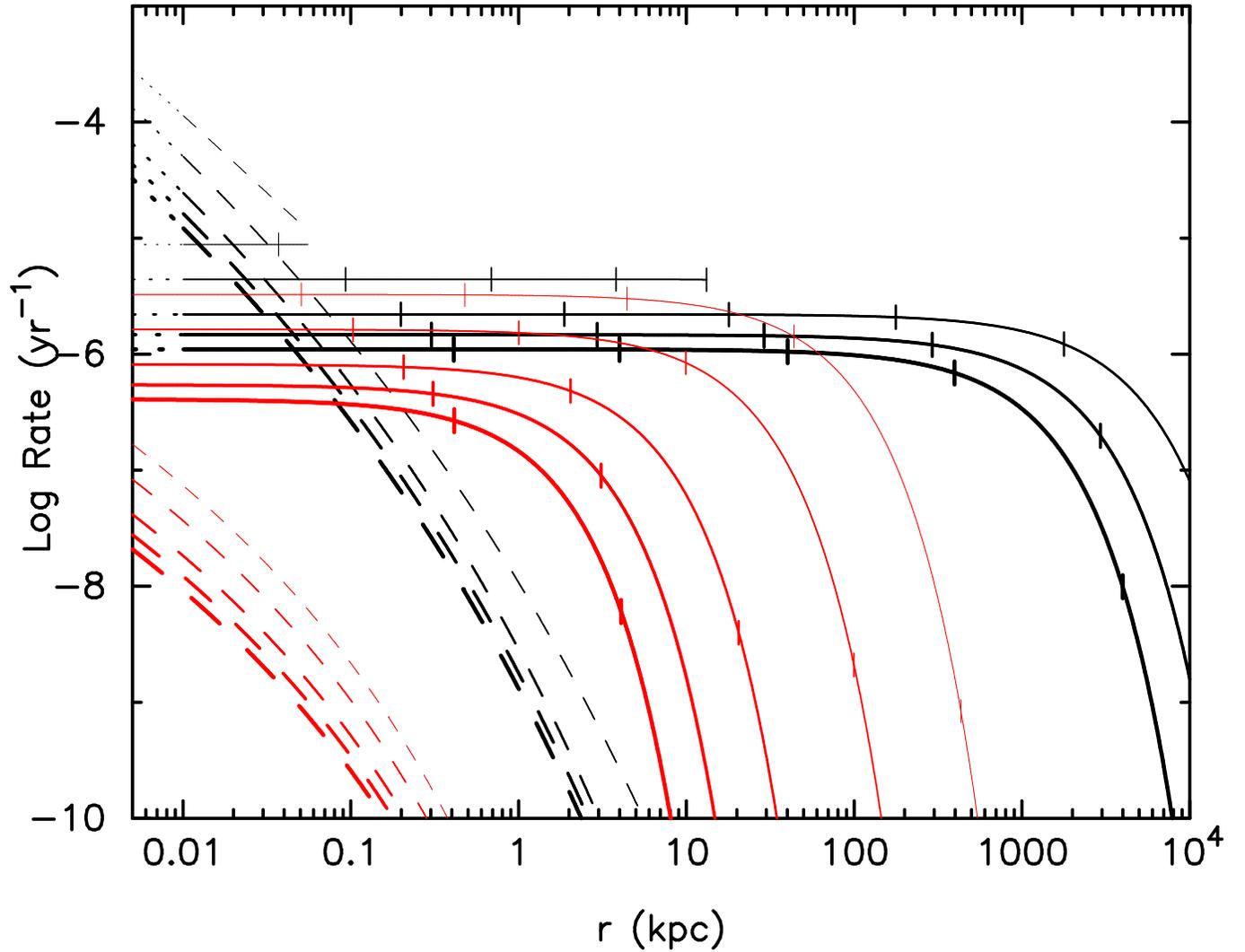}
\caption{Tidal disruption rate as a function of distance
travelled for SMBHs ejected from the centers of two galaxy models. 
{\it Model 1 (black lines):} $M_{\rm gal}=4.5\times 10^{10}\msun,
\mh=3\times 10^7\msun, {\rm Sersic~index~} n=4.0, R_e=1 {\rm kpc}, \rinfl=10 {\rm pc}$.
{\it Model 2 (red lines):} $M_{\rm gal}=1.5\times 10^{9}\msun,
\mh=1\times 10^6\msun, n=2.5, R_e=0.3 {\rm kpc}, \rinfl=3 {\rm pc}$.
Line width indicates kick velocity: $\vk=500 $\kms (thinnest),
$1000,2000,3000,4000$ \kms (thickest).
Solid lines show $\dot N$ for bound stars, dashed lines for
unbound stars.
Lines terminate at the right where the SMBH reaches apocenter;
distances $r\le\rinfl$ are indicated via dotted lines.
The falloff in the bound rates at large radii is due to 
depletion of the bound population. The tick marks indicate time and are
at log (t/yr) = 5,6,7,8,9.
}
\end{figure}

\end{document}